\documentclass[natbib]{svjour3}                     
\smartqed  
\usepackage[pdftex]{graphicx}
 \usepackage{aps-bibstyle}  
\begin{document}

\title{Fast variability from black-hole binaries
}


\author{Tomaso M. Belloni         \and
            Luigi Stella 
}


\institute{T.M. Belloni \at
              INAF - Osservatorio Astronomico di Brera, Via E. Bianchi 46, I-23807 Merate, Italy \\
              Tel.: +39-039-5971026\\
              Fax: +39-039-5971001\\
              \email{tomaso.belloni@brera.inaf.it}           
           \and
           L. Stella \at
              INAF - Osservatorio Astronomico di Roma, Via di Frascati, 33, 00040 Monte Porzio Catone, Italy\\
              Tel.: +39-06-94286436\\
              Fax: +39-06-9447243\\
              \email{luigi.stella@oa-roma.inaf.it}
}

\date{Received: date / Accepted: date}

\maketitle

\begin{abstract}
Currently available information on fast variability of the X-ray emission from accreting collapsed objects constitutes a complex phenomenology which is difficult to interpret.  We review the current observational standpoint for black-hole binaries and survey models that have been proposed to interpret it. Despite the  complex structure of the accretion flow, key observational 
diagnostics have been identified which can provide direct access to the dynamics of matter motions in the close vicinity 
of black holes and thus to the some of fundamental properties of curved spacetimes, where 
strong-field general relativistic effects can be observed. 


\end{abstract}

\section{Introduction}
\label{intro}

The first power density spectra (PDS) from the black-hole binary (BHB) Cyg X-1 were obtained
with the Uhuru satellite and it became clear that no obvious periodicity was present, but the data were
consistent with a noise process (Terrell 1972). For two decades, the dearth of known sources
and the small effective area of X-ray instruments allowed only the study of the PDS of Cyg X-1, which 
features strong continuous noise (10-40\% fractional rms), with a few breaks which allow to identify
characteristic time scales (Nolan et al. 1981, Belloni \& Hasinger 1990). Only with the Ginga satellite, which included a large-area 
instrument together with an all-sky monitor to detect transient sources, quasi-periodic oscillations (QPOs) at frequencies
between 0.01 and 10 Hz were found in a few systems (see Tanaka \& Lewin 1996 for a review). The
launch of the Rossi X-ray Timing Explorer (RXTE) increased dramatically the number of known sources, their coverage
and the detection sensitivity, allowing classification of these low-frequency QPOs into different flavours (see Motta et al. 2011), as well as the discovery of weak signals at higher frequencies (see e.g. Belloni, Sanna \& M\'endez 2012 and references therein). Here, we briefly summarise the basic knowledge deriving from this observational body and discuss its relevance for accretion theory and general relativity (GR).
\section{Accretion and General Relativity}

Early in the development of models of accretion disks around black holes it was realised that disk inhomogeneities 
orbiting in the innermost regions of the disk, where the bulk of the energy is released, can give rise to 
fast X-ray variability. If the lifetime and radial drift timescale of such inhomogeneities is longer than 
their orbital period, a quasi periodic signal ensues that can provide key information 
on the properties of matter motion and light deflection in the strong field regime of general relativity.
For instance, back in 1972 Sunyeav discussed the possibility to identify black holes and ``discriminate 
between the cases where a Schwarzschild or Kerr metric prevails'', based on the fastest QPO signals arising 
from a disk at its innermost stable circular orbit (ISCO) (Sunyaev 1972; see also Shakura \& Sunyeav 1973,
Novikov \& Thorne 1973). Stoeger (1980) went even further and calculated the characteristics
of quasi-periodic signals produced by matter spiralling inwards in the unstable region between the inner disk 
boundary (considered to be at the ISCO radius) and the black horizon. These theoretical works as well as others 
were published well before the first detections of QPOs in accreting black holes 
and neutron stars and provided  in many cases the framework within which QPOs were interpreted and models further 
developed.

\begin{figure*}
\includegraphics[width=1.0\textwidth]{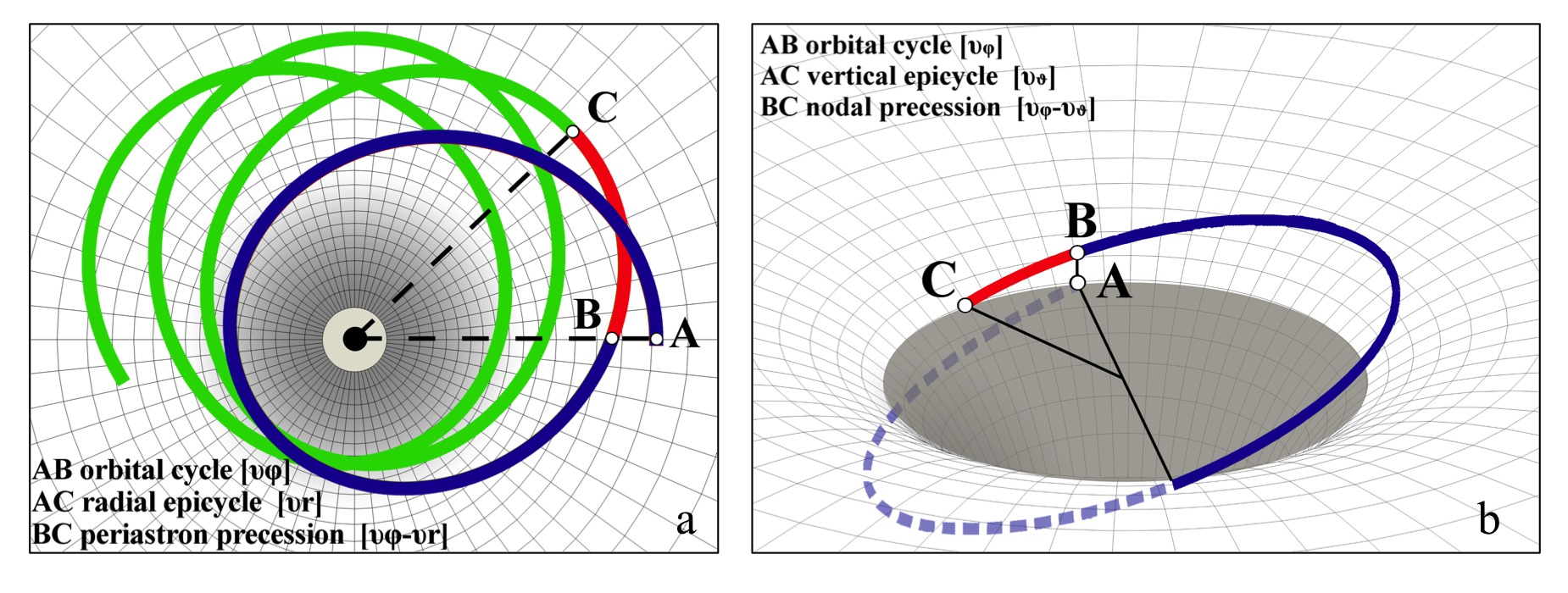}
\caption{Example of an eccentric and tilted orbit around a Kerr black, as seen face on (left panel), and from a $\sim 60$~deg inclination angle (right panel). Cycles are represented for each of the three different fundamental frequencies of motion: azimuthal (aka orbital), and radial epicyclic and vertical epicyclic. The way in which the orbit undergoes periastron and nodal precession is also shown. Embedding diagrams are plotted to help visualize the perspective.}
\label{fig:orbits}      
\end{figure*}

\begin{figure*}
\includegraphics[width=1.0\textwidth]{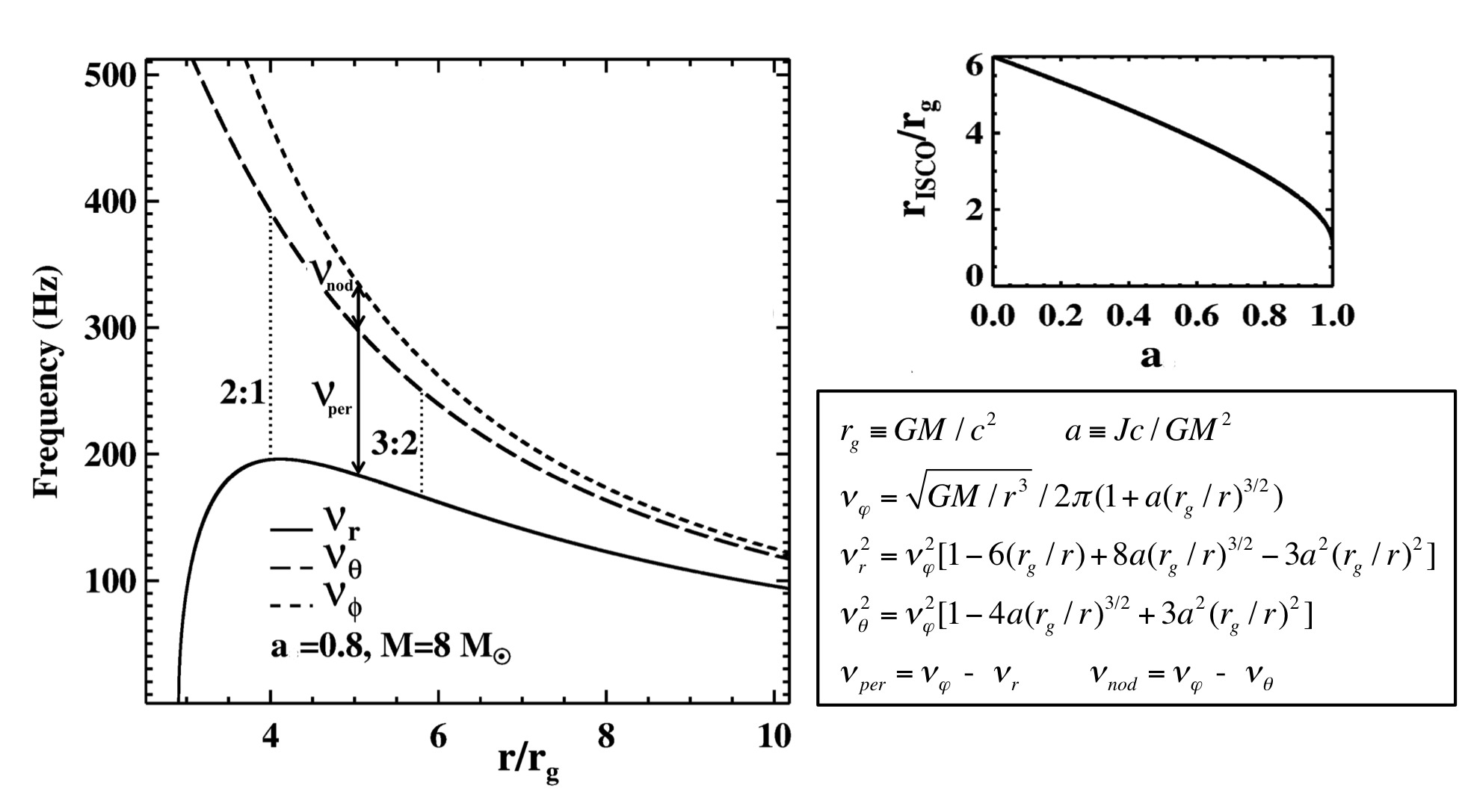}
\caption{Left panel: Fundamental frequencies of motion for an infinitesimally eccentric and tilted orbit 
around a Kerr black hole with $M= 8$~M$_\odot$ and $a=0.8$, as a function of radius. In contrast to the other two frequencies, the radial epicyclic frequency decreases with radius close to the compact object, reaching zero at the ISCO radius. The dotted lines show the radii at which the two epicyclic frequencies are in a 2:1 and 3:2 ratio. The double arrowed segments show 
the periastron and 
nodal precession frequencies at a radius of 5M. Right top panel: ISCO radius as a function of the 
Kerr spin parameter $a$ for corotating orbits. Right bottom panel: formulae for the fundamental frequencies in the Kerr metric.}
\label{fig:freqs}      
\end{figure*}

Based on the discovery of multiple, simultaneous QPOs whose frequency varies in some cases in a correlated fashion 
(especially neutron star systems, see van der Klis 2006 for a review) 
different scenarios were proposed in which different quasi-periodic phenomena can take place at the 
same time. Amongst these are models that involve the fundamental frequencies of bound motion around 
a collapsed object, namely the azimuthal (aka orbital) frequency $\nu_\phi$, the radial epicyclic frequency 
$\nu_r$ and the vertical epicyclic frequency $\nu_\theta$ (for a sketch of the relevant motions see 
Fig. \ref{fig:orbits}). The expressions for these frequencies as a function of the radius $r$ of slightly eccentric, 
slightly tilted orbits close to the equatorial plane of a Kerr black hole are given in Fig. \ref{fig:freqs}, where
$r_g = GM/c^2$ is the gravitational radius, $a = Jc/GM^2$ the spin parameter and $J$ the specific angular momentum 
of the black hole. The behaviour of the fundamental frequencies of matter motion in the
strong field gravity regime that characterizes the inner region of accretion flows towards
black holes (see Fig.~\ref{fig:freqs} for an example), besides being  considerably different from that 
of a point mass in Newtonian gravity (for which all 
three frequencies are identical), cannot be approximated by weak field expansions of General Relativity (GR).
For instance, the radial epicyclic frequency increases from zero at the ISCO, to a broad maximum at 
a $\sim 40$\% larger radii and then decreases as $r^{-3/2}$: this behavior is 
characteristic of the strong gravitational field close to a collapsed object. 
Therefore if QPOs can be associated unambiguously to fundamental frequencies of motion 
(or combinations or functions of them), they hold the potential to verify some key predictions 
of GR in the strong field regime, and measure the mass and spin of black holes.  
We note that the timescale corresponding to these frequencies
are usually faster than viscous and thermal timescales of the disk (e.g. Pringle 1981). 

\label{sec:theory}
\section{Fast variability and source states}
\label{sec:observations}

The low-frequency ($0.01-30$ Hz) fast variability from BHBs can take different forms and can range from being barely detectable to very strong. It is clear that what is observed is strongly connected to the state of the source, i.e. to the shape of the energy spectrum. 
For a definition of states, see Belloni (2010) and Belloni, Motta \& Mu\~noz-Darias (2011).
We divide here the discussion in two parts: the Low Hard State (LHS) and High Soft State (HSS) with their noise components, and the intermediate states with their QPOs.

\subsection{Hard and soft states: noise}
\label{sec:hardsoft}

\begin{figure*}
  \includegraphics[width=1.0\textwidth]{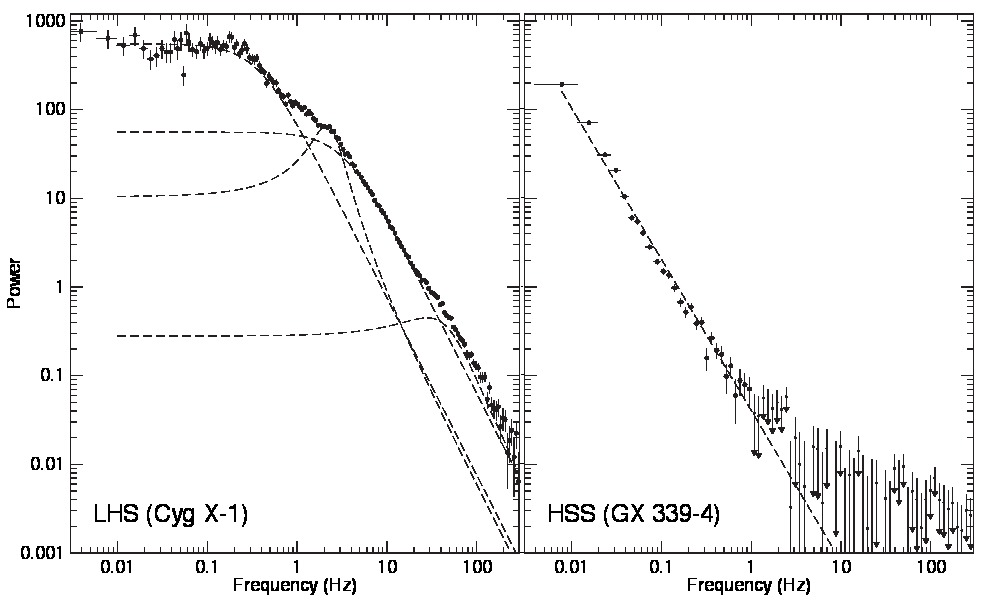}
\caption{Left: Representative PDS from the LHS obtained with the RossiXTE satellite (Cyg X-1). The dashed lines indicate
separate Lorentzian components used for the fit. Right: PDS from the HSS (GX 339-4). The dashed line indicate a power law fit.}
\label{fig:pds1}       
\end{figure*}

For a full description of the energy spectra that characterise the LHS and HSS see Gilfanov (2010). 
In the LHS, the X-ray emission is very noisy, with fractional rms values as high as 40-50\%. The PDS shows that this variability is in form of band-limited noise, with usually no peaked components or QPOs (see Fig. \ref{fig:pds1}, left panel). The shape
of the PDS is not compatible with a simple model and is usually fitted with a combination of Lorentzian components (Belloni, Psaltis \& van der Klis 2002). These components provide characteristic frequencies (see Belloni, Psaltis \& van der Klis 2002 for a detailed description), but since they are very broad the determination of a characteristic frequency is not unique as in the case of a narrow peak and are therefore model dependent. Four main components have been identified: a low-frequency one providing the flat-top part ($L_b$), a peaked (sometimes QPO-like) component ($L_q$) and two broad Lorentzians at higher frequencies ($L_\ell$ and $L_u$).
In a black-hole transient, the LHS is observed at the start and end of the outburst. In the early phases, as source flux (and accretion rate) increases, the total rms decreases slightly (see e.g. Mu\~noz-Darias, Motta \& Belloni 2011), all characteristic frequencies increase and the energy spectrum steepens. The reverse takes place at the end of the outburst. In Cyg X-1, a persistent source, there is only one of these components whose frequency does not appear to vary (Pottschmidt et al. 2003).

It is difficult to associate the observed characteristic frequencies to physical frequencies either from the accretion flow or related to General Relativity. This is possible only when considering the intermediate states, to which the observed correlations extend (see below). One interesting correlation is the one that links the break frequency of the lowest component (the flat-top break $\nu_b$) and the flat top fractional rms value (Belloni \& Hasinger 1990; Belloni, Psaltis \& van der Klis 2002).

Both the linear relation observed over a broad range of frequencies between rms variability and source intensity, and the log-normal distribution of count rates indicate that the strong variability observed in the LHS is not consistent with being a linear process. (see Heil, Vaughan \& Uttley 2012 and references therein). Propagating fluctuation models have been proposed to incorporate these observables (Lyubarskii 1997).

The LHS noise is slightly stronger at lower energies. The fractional rms as a function of energy (the``rms spectrum") is flat or decreases by a few \% from 2 to 20 keV (see Belloni, Motta \& Mu\~noz-Darias 2011 and references therein). The rms spectrum has been interpreted in the framework of Comptonization models (Gierli\'nski \& Zdziarski 2005). Recent studies with XMM-Newton have shown that the soft thermal disk emission present below 1 keV leads the hard emission at time scales longer than one second, while it lags at shorter time scales, consistent with propagating models modified by disk heating at short time scales (Uttley et al. 2011).
Finally, in the LHS the variability at high energies is observed to lag that at lower frequencies (see e.g. Nowak et al. 1999), consistent with a Comptonization origin for the emission (Gilfanov 2010).


In the HSS, variability is limited to a few \% and the typical PDS features a power law component (see Fig. \ref{fig:pds1}, right panel). Weak QPOs have been detected in some sources (see e.g. Motta et al. 2012), identified as high-frequency extension of type C QPOs (see below). The energy spectrum is dominated by the thermal disk component (See Gilfanov 2010), which is not much variable (but see Uttley \& Casella, this book). It is possible that most or all of the observed variability is associated to the faint hard component which is usually observed (see Grove et al. 1998).

\subsection{Intermediate states: quasi-periodic oscillations}
\label{sec:intermediate}

\begin{figure*}
  \includegraphics[width=1.0\textwidth]{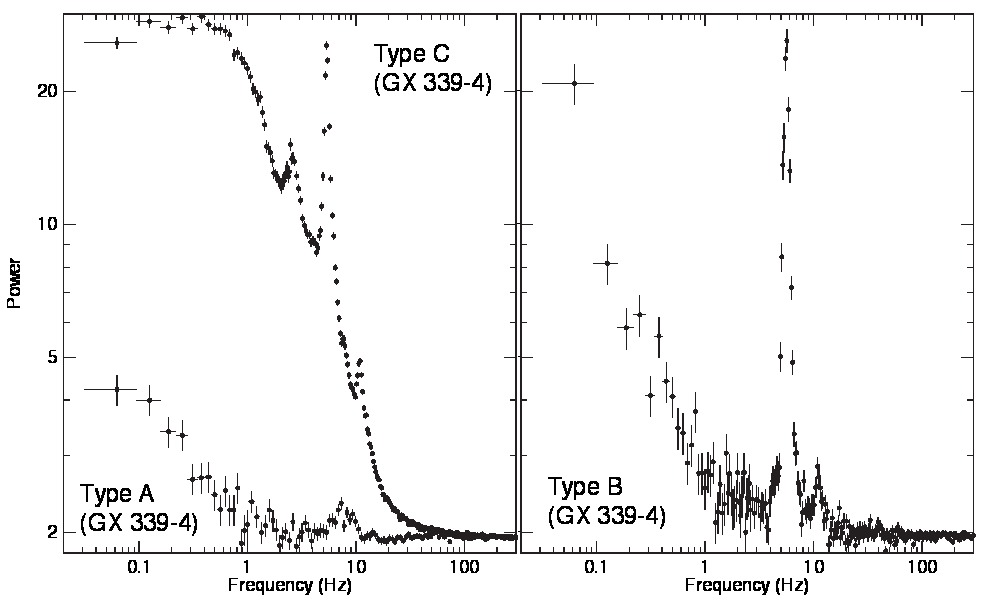}
\caption{Representative PDS if with a type C and type A QPO (left), and with a type B QPO (right). All three PDS are from RossiXTE observations of GX 339-4 (see Motta et al. 2011).}
\label{fig:pds2} 
\end{figure*}


In most cases, considerable time during an outburst is spent in the LHS and HSS (Fender \& Belloni 2012). It is however
during the transition between these states that the most prominent features in the fast time variability are observed. Two separate intermediate states have been identified, called Soft Intermediate State (SIMS) and Hard Intermediate State (HIMS). The boundary between these to states is defined by sharp changes in the variability.

\begin{itemize}

\item The HIMS is when the most common type of LFQPO is observed, called 'type C QPO' (see Wijnands, Homan \& van der Klis 1999; Motta et al. 2011 and references therein). The most important characteristics of this oscillation is that it is observed to vary over a relatively large range of frequencies, roughly from 0.01 to 30 Hz (see Fig. \ref{fig:pds2} for an example). These oscillations are observed in a large number of sources (Belloni 2010, Remillard \& McClintock 2006) and were the first oscillations discovered from BHBs. Their typical quality factor Q (defined as the centroid frequency divided by the FWHM of the peak in the PDS) is around 10 (see e.g. Casella, Belloni \& Stella 2005; Rao et al. 2010) and their fractional rms variability of the order of 3--15\%.  Often it appears with one or two overtones and at times with a sub-harmonic (see \ref{fig:pds2}). It is always associated to a band limited noise and its frequency is anti correlated with the total broad-band fractional rms variability (see lower left panel in Fig. \ref{fig:sara}, from Motta et al. 2011). The total fractional rms (QPO peaks plus noise components) of observations with type C QPOs is between 10 and 30\% (Mu\~noz-Darias et al. 2011).

The energy spectrum of type C QPOs is hard, like that of all BHB QPOs with the rms vs. energy increasing and flattening above 10 keV (see Casella et al. 2004), with some evidence of a decrease at higher energies (Rodriguez et al. 2004). Despite the fact that the QPO is present (and more intense) at high energies where the accretion disk does not contribute, the centroid frequency is correlated with the soft flux from the accretion disk (see Markwardt, Swank \& Taam 1999; Motta et al. 2011). Type C QPOs are often also observed in the LHS, at lower frequencies. In those cases, the frequency evolution from LHS to HIMS is continuous.

The large span in centroid frequency of type C QPOs is important to identify their origin. Wijnands \& van der Klis (1999) discovered  a strong correlation between their frequency and the break frequency of the main noise component (around 1Hz in the left panel of Fig. \ref{fig:pds1}). This correlation is also valid for neutron star binaries when one considers the so-called Horizontal Branch Oscillations (HBO), quasi-periodic peaks which have been associated to type C QPO (van der Klis 2006; Casella, Belloni \& Stella (2005).

\begin{figure*}
\begin{tabular}{@{}cc@{}}
  \includegraphics[width=0.49\textwidth]{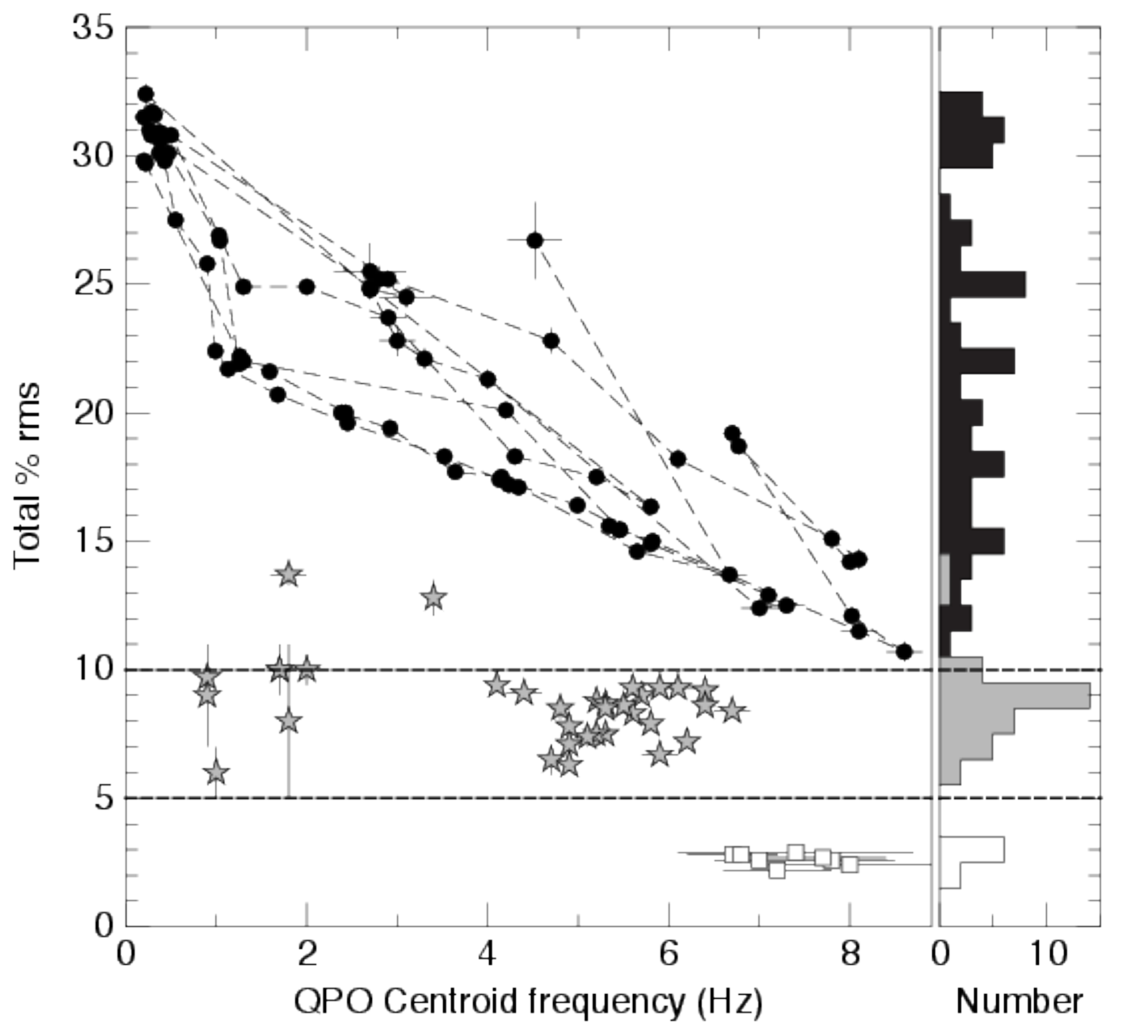} & \includegraphics[width=0.49\textwidth]{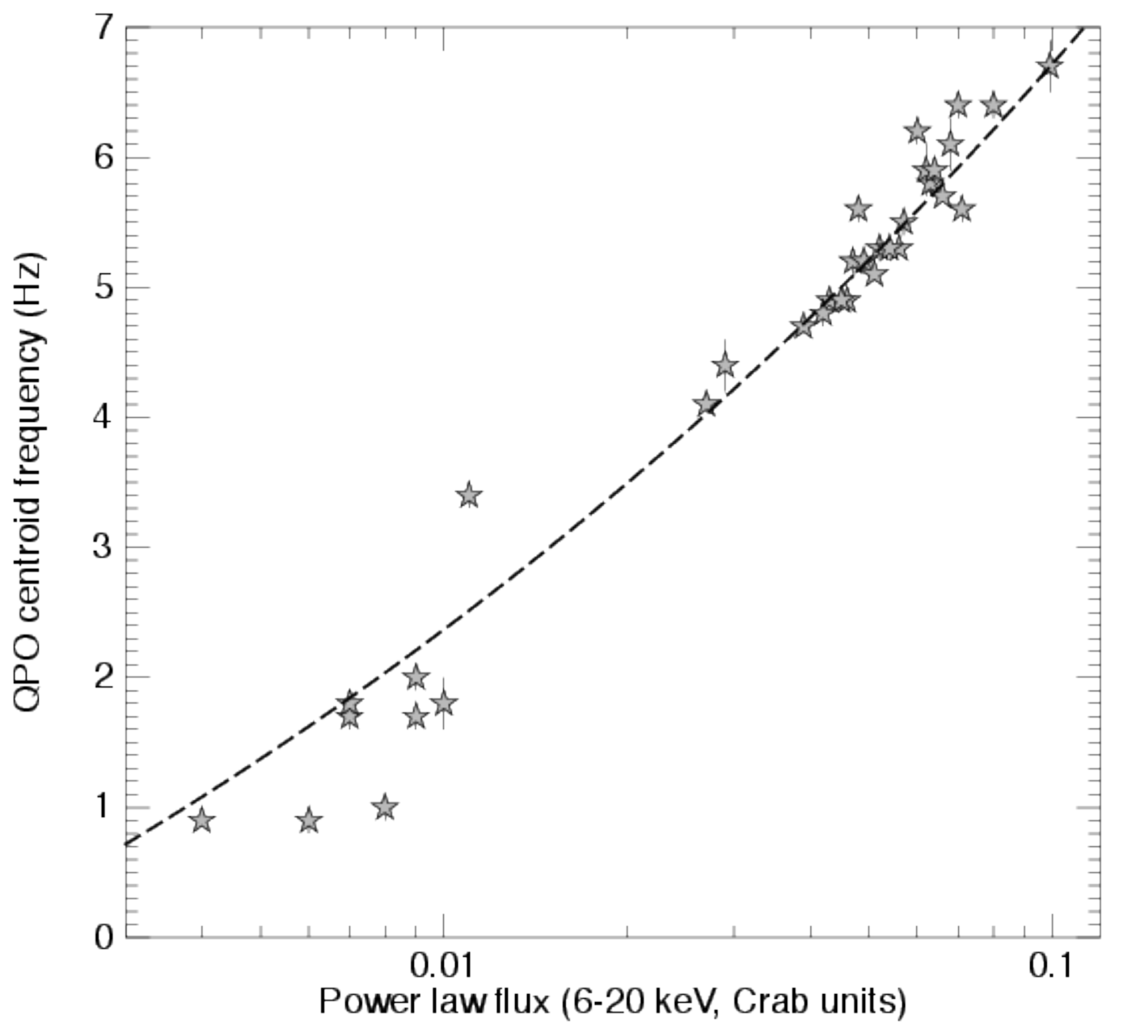}\\
  \end{tabular}
\caption{Left: total fractional rms vs. QPO centroid frequency for all RXTE QPOs of GX 339-4, with the rms marginal distribution on the right side. The three types of QPOs (type A: white squares; type B: grey stars; type C: black circles) are clearly separated. 
Right: correlation between centroid frequency and power law flux ($<$6 keV) for type B QPOs in the same GX 339-4 data
(both panels adapted from Motta et al. 2011).
}
\label{fig:sara} 
\end{figure*}

\item The SIMS, a short-lived state characterised by a softer energy spectrum than the HIMS, is defined in terms of sharp changes of properties of time variability. The band-limited noise is replaced by a weaker power-law noise component, and two alternative different types of QPO, called type A and type B,  replace the type C QPO. These three types of QPO have been shown not to be the same signal, based on the fast transition between them (see e.g. Nespoli et al. 2003; Casella et al. 2004; Belloni et al. 2005) and in particular on the simultaneous detection of a type B and a type C (Motta et al. 2012). An example of each type of QPO can be seen in Fig. \ref{fig:pds2}). Type A  QPOs have centroid frequencies 6.5-8 Hz, are broad (quality factor Q$\sim$1-3) and weak (fractional rms less than 5\%). Type B QPOs have centroid frequencies 0.8-6.4 Hz, are narrow (Q$>$6) and have a 5-10\% fractional rms, in addition to appearing often together with an overtone and a sub-harmonic. In the left panel of Fig. \ref{fig:sara} one can see that the three types of QPO are well separated as a function of the {\it total} integrated fractional rms in the PDS (noise+QPOs, see also Mu\~noz-Darias, Motta \& Belloni 2011). The centroid frequency of type B QPOs shows a strong correlation with the flux from the hard component, as measured above 6 keV (right panel in \ref{fig:sara}).

Not much is known about type A QPOs, given their relative weakness and large FWHM, which prevent their detection in most cases. They are associated to a softer spectrum than type B QPOs. Both QPOs are observed in the SIMS, which is a short lived and transient state, possibly associated to the ejection of relativistic jets (see Fender, Homan \& Belloni 2009 and references therein). They are much rarer than type C QPOs and span a smaller range in frequency. 

\end{itemize}

A correspondence between the three flavours of LFQPO in BHBs and the three types of QPOs in neutron star binaries (normal branch oscillations: NBO; horizontal branch oscillations: HBO and flaring branch oscillations: FBO) has been suggested (Casella, Belloni \& Stella 2005; van der Klis 2006). The similarities between type C QPOs and HBOs are very strong (van der Klis 2006) and the existence of a general correlation such as that described above strengthens this hypothesis (Wijnands \& van der Klis 1999).

\section{Variability at high-frequencies}
\label{sec:hfqpo}
The large collecting area of the RXTE/PCA, which allowed the discovery of kHz QPOs in neutron star binaries
(see van der Klis 2006), also opened a window onto high-frequency phenomena in BHBs. The first observations
of the very bright system GRS 1915+105 led to the discovery of a transient oscillation at $\sim$67 Hz (Morgan,
Remillard \& Greiner 1997). Since then, sixteen years of RXTE observations have yielded only a handful of  
detections in other sources, although GRS 1915+105 seems to be an exception (see Belloni, Sanna \& M\'endez 2012; Altamirano \& Belloni 2012; Belloni \& Altamirano 2013a,b, and references therein).

\begin{table}
\caption{Narrow HFQPO features from BHBs. The columns are: source name, number of detected peaks, flag for simultaneity for the multiple peaks, approximate centroid frequencies. Question marks are discussed in the text}
\label{tab:1}       
\begin{tabular}{lccc}
\hline\noalign{\smallskip}
Source & N$_{peaks}$ & Simultaneity & Frequency (Hz)  \\
\noalign{\smallskip}\hline\noalign{\smallskip}
GRO J1655-40   & 2 & Y & $\sim$300, $\sim$400 \\
XTE J1550-564  & 2 & N & $\sim$180, $\sim$280 \\
XTE J1650-500  & 1 & -  & $\sim$250 \\
H 1743-322        & 2 & N & $\sim$160, $\sim$240 \\
IGR 17091-3624& 1 & -   & $\sim$66 \\
GRS 1915+105  & 4?  & Y & $\sim$27, $\sim$34, $\sim$41, $\sim$67 \\
\noalign{\smallskip}\hline
\end{tabular}
\end{table}

Some of the reported detections were at low statistical significance, while others involved components too broad to be defined as QPO and difficult to assess due to the presence of counting noise. The remaining ones belong to six sources (see Tab. \ref{tab:1}). Their properties can be summarized as follows:

\begin{itemize}

\item They are observed only in observations at high flux/accretion rate.  This is obviously at least partly due to a selection effect, but not all high-flux observations lead to the detection of a HFQPO, all else being equal, indicating that the properties of these oscillations can vary substantially even when all other observables do not change.

\begin{figure*}
\begin{tabular}{@{}cc@{}}
  \includegraphics[width=0.49\textwidth]{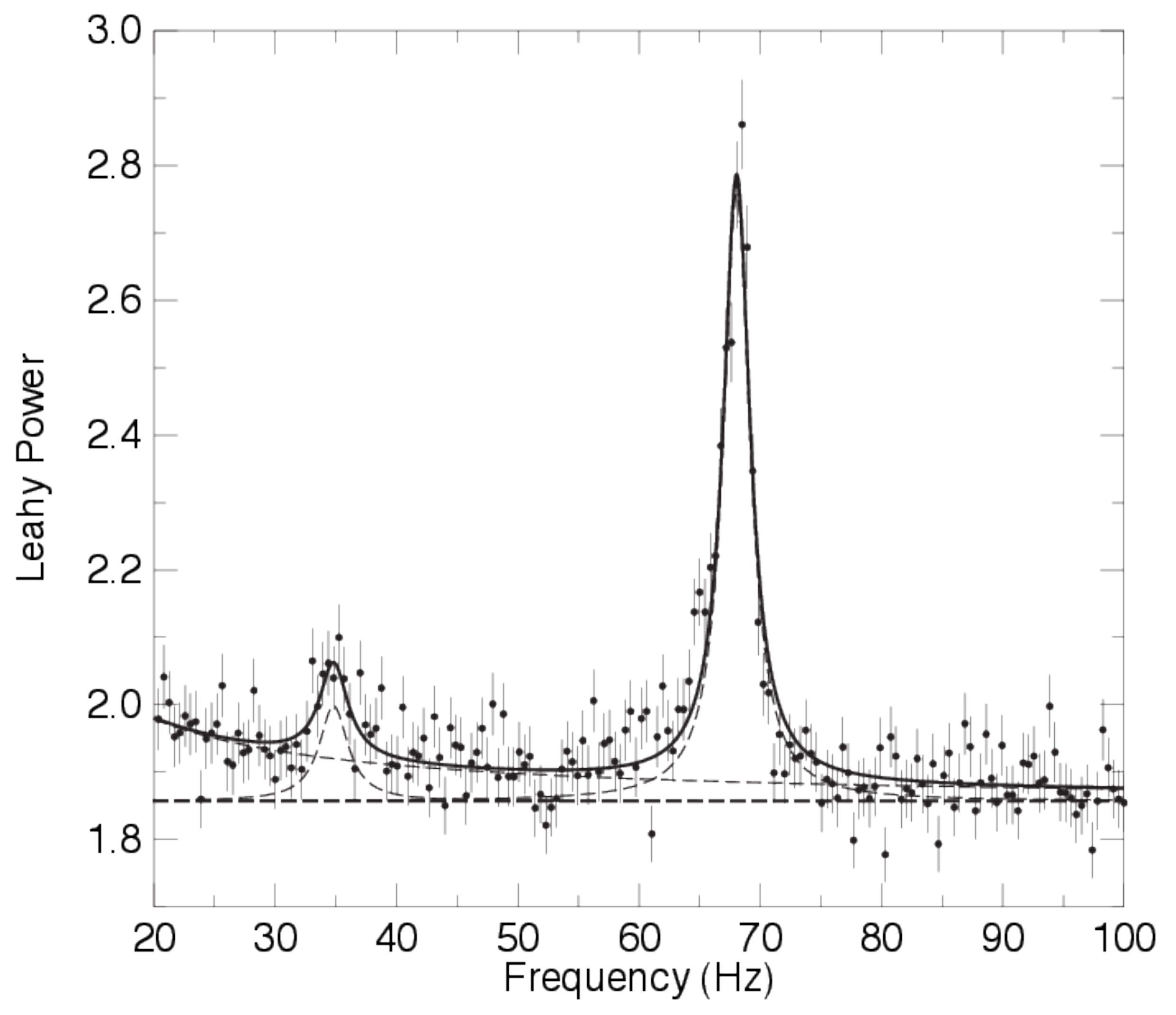} & \includegraphics[width=0.49\textwidth]{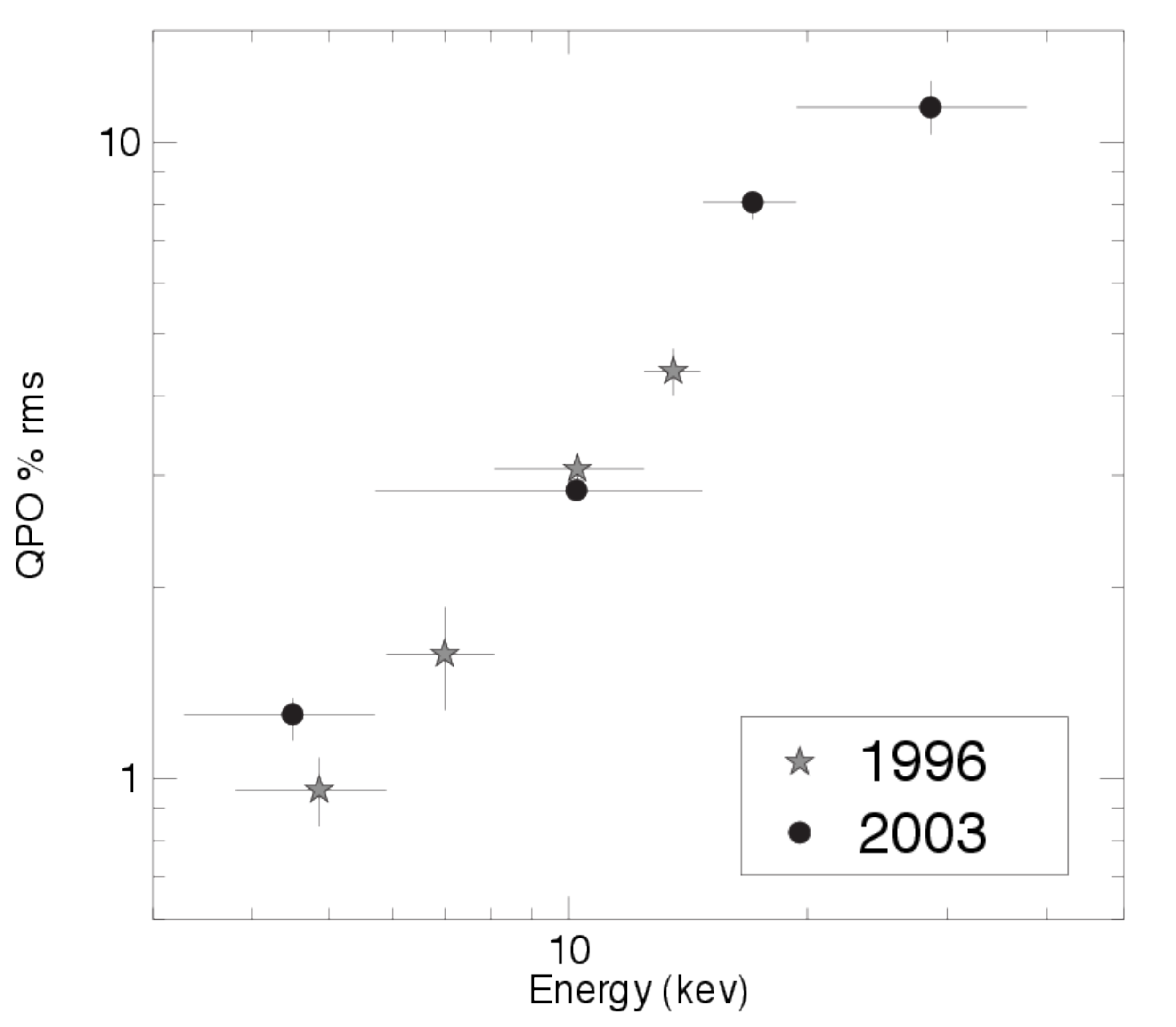}\\
  \end{tabular}
\caption{Left: two HFQPO at 34 Hz and 68 Hz in a PDS of RXTE data of GRS 1915+105 (Belloni \& Altamirano 2013b). 
Right: fractional rms as a function of energy for the 67 Hz QPO in GRS 1915+105 at two different epochs. From Belloni and Altamirano (2013a).
}
\label{fig:pds_sara} 
\end{figure*}

\item They are observed as single or double peaks. From Tab. \ref{tab:1}, two sources have shown a single peak: XTE J1650-500 (Homan et al. 2003) and IGR J17091-3624 (Altamirano \& Belloni 2012). For the others, GRS J1655-40 showed two clear simultaneous peaks (Strohmayer 2001a; Belloni, Sanna \& M\'endez 2012; Motta et al. 2014a). In XTE J1550-564, the two detected peaks (Remillard et al. 2002; Belloni, Sanna \& M\'endez 2012) have been detected simultaneously, but the lower one with a 2.3$\sigma$ significance after taking into account number of trials (Miller et al. 2001). M\'endez et al. (2013), on the basis of their phase lags, suggested that the two detected peaks might be the same physical signal at two different frequencies. For H 1743-322, there is a weak detection of a second simultaneous peak (Homan et al. 2005), but two significant peaks have been detected, although not simultaneously (Remillard et al. 2006).

\item While type C QPOs at low frequencies are rather common and define the presence of the HIMS, not only HFQPOs are very rarely detected with RXTE, but their detection is almost completely anticorrelated with the presence of type C QPOs. The only exception to this exclusion rule is GRO J1655-40 (see below). Often, HFQPOs are associated to type B QPOs or to the source being in the ``anomalous state'' (Belloni 2010; Belloni, Sanna \& M\'endez 2012).

\item  A different case is that of GRS 1915+105, the first source where a HFQPO was discovered (Morgan et al. 1997). From the large archive of RXTE observations of this peculiar system (see Fender \& Belloni 2004 for a review), a systematic analysis led to the detection of 51 HFQPOs, 48 of which at a centroid frequency between 63 and 71 Hz (Belloni \& Altamirano 2013). This indicate that this range of frequencies is in some way special for this system. All detections corresponded to a very limited range in spectral parameters, as measured through hardness ratios. With a special time selection, Belloni, M\'endez \& S\'anchez-Fern\'andez (2001) found a single 27 Hz QPO. Strohmayer (2001b), averaging observations, discovered a 41 Hz HFQPO simultaneous to the 67 Hz one. Finally, Belloni \& Altamirano (2013) discovered a 34 Hz peak simultaneous with a 68 Hz one (Fig. \ref{fig:pds_sara}, left panel). This could be the same as the 41 Hz one, in which case only one frequency would have changed, or a sub-harmonic of the upper one. The most recent HFQPO discovered, in IGR J17091Ð3624, is consistent with the average frequency of the 67 Hz QPO in GRS 1915+105 (Altamirano \& Belloni 2012).

\item Typical fractional rms for HFQPOs are 0.5-6\% (Belloni, Sanna \& M\'endez 2012; Belloni \& Altamirano 2013). The fractional rms increases steeply with energy, in the case of GRS 1915+105 reaching more than 19\% at 20-40 keV (see right panel in Fig. \ref{fig:pds_sara}; Morgan et al. 1997; Belloni \& Altamirano 2013). Quality factors ($Q$, the ratio between centroid frequency and FWHM of the QPO peak) are around 5 for the lower peak and 10 for the upper. In GRS 1915+105, a typical $Q$ of $\sim$20 is observed, but values as low as 5 and as high as 30 are observed.

\item Time lags of HFQPOs have been studied for four sources (M\'endez et al. 2013). The lag spectra of the 67 Hz QPO in GRS 1915+105 and IGR J17091Ð3624 and of the 450 Hz QPO in GRO J1655-40 are hard (hard photons variations lag soft photons variations), while those of the 35 Hz QPO in GRS 1915+105 are soft. The 300 Hz QPO in GRO J1655-40 and both HFQPOs in XTE J1550-564 are consistent with zero. The fact that the lag spectrum of both QPOs in the latter source are the same suggests that the 180 Hz and 280 Hz QPOs in this source might represent the same physical oscillation observed at two different frequencies.

\item For three sources, GRO J1655-40, XTE J1550-564 and XTE J1743-322, the two observed frequencies are close to being in a 3:2 ratio (Strohmayer 2001a, Remillard et al. 2002, Remillard et al. 2006), which has led to a specific model for their physical origin (see Sect.~5.2). For GRS 1915+105 the 67 Hz and 41 Hz QPOs, observed simultaneously, are roughly in 5:3 ratio. The 27 Hz would correspond to 2 in this sequence.
	
\end{itemize}

\begin{figure*}
\begin{tabular}{@{}cc@{}}
  \includegraphics[width=0.49\textwidth]{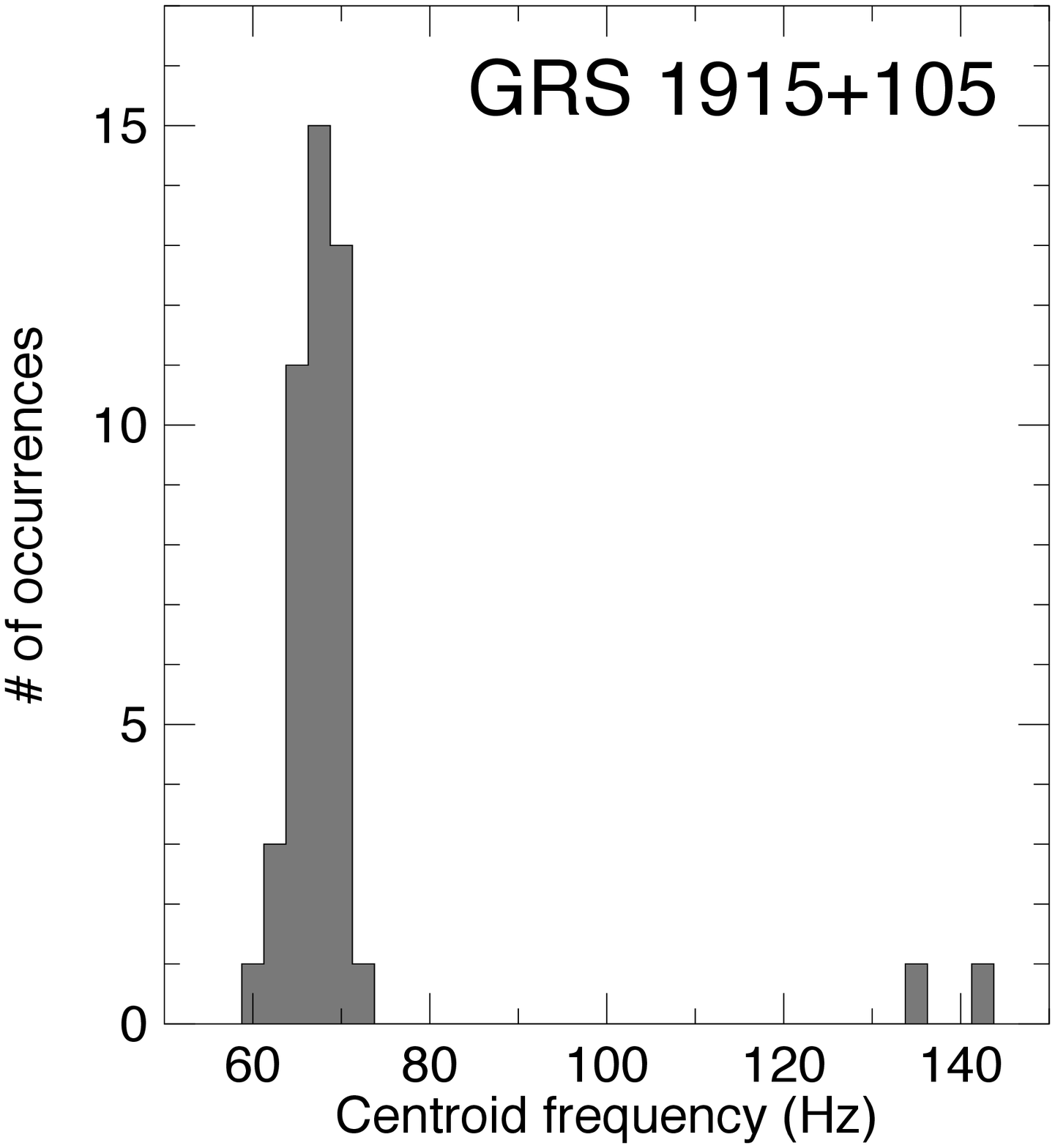} & \includegraphics[width=0.49\textwidth]{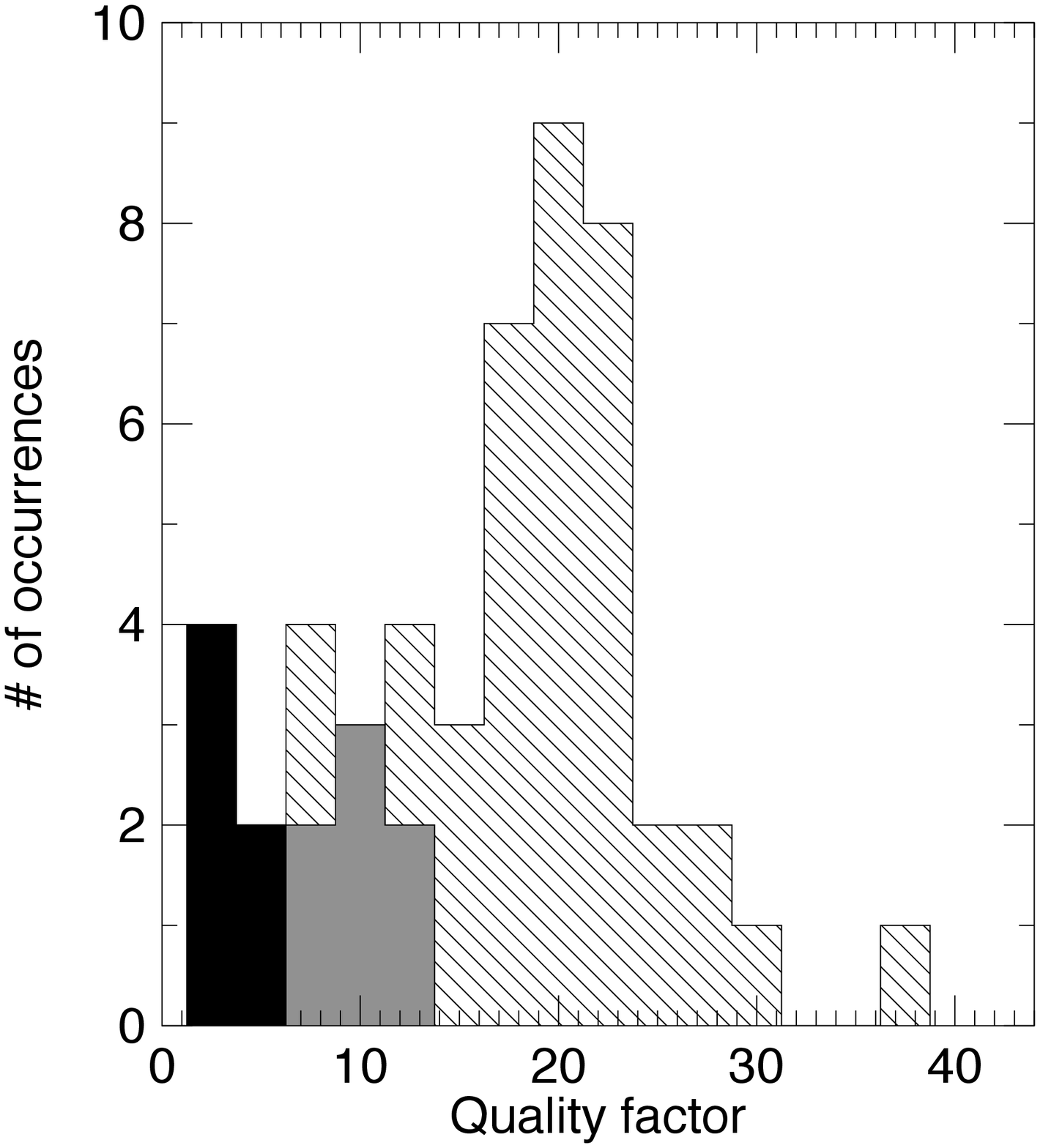}\\
  \end{tabular}
\caption{Left: distribution of centroid frequencies of the HFQPOs detected in the full RXTE archive of observations of GRS 1915+105 (adapted from Belloni \& Altamirano 2013a). Right: distribution of quality factors for the HFQPOs of GRS 1915+105 (dashed histogram, from Belloni \& Altamirano 2013a), simultaneous pairs of HFQPOs in other sources (black: upper peak, tray: lower peak; from Belloni, Sanna \& M\'endez 2012, Remillard et al. 2006 for H 1743-322). The single HFQPO in XTE J1650-500 (Homan et al. 2003) and the 34 Hz QPO in GRS 1915+105 (Belloni \& Altamirano 2013b) have a $Q$ of $\sim$5.
}
\label{fig:histo_hfqpo} 
\end{figure*}

As for LFQPOs, it is interesting to compare the properties of HFQPOs with those of high-frequency (kHz) QPOs in neutron star binaries (see van der Klis 2006 for a review). kHz QPOs also appear often in pairs and at frequencies generally higher than those of HFQPOs. However, they are much stronger signals, with a large number of detections in all bright systems, they are
observed to change their frequency while keeping a roughly (but not exactly) similar frequency difference, and they correlate with other properties of the systems, including a possible relation with the neutron star spin frequency (but see M\'endez \& Belloni 2007).
The phenomenology for NS kHz QPOs is complex and rather well defined, thanks to the large number of available detections.

Psaltis, Belloni \& van der Klis (1999; PBK) found that the known correlation existing in NS systems between HBO oscillations and the lower of the two kHz QPOs could be extended to lower frequencies if the kHz QPO frequency was replaced by the characteristic frequency of one of the broad components found in the PDS at low luminosities. This correlation was found to be valid also for black hole systems, who show the same high-frequency broad component simultaneous to a type C QPO (see the left panel of Fig. \ref{fig:pds1}). This constitutes a strong link between the two classes of sources and suggests that the frequencies are associated to the same physical processes (see Sect.~5).

\section{Interpretation}
\label{sec:interpretation}
Early modeling of the fast variability properties of black hole X-ray binaries, concentrated 
on exponential shot noise and other aperiodic processes which can interpret the continuum
power spectral components that were known at the time (Terrell 1972; 
Nolan et al. 1981; Belloni \& Hasinger 1990). 
Then in the mid-eighties $5-50$~Hz QPOs with $rms$ amplitudes of several percents 
were discovered in the X-ray flux of neutron star low mass X-ray binaries. 
Being these QPOs tens to hundreds times slower than the orbital frequencies of matter 
in the innermost disk, interpretations were devised in which 
relatively low frequency signals could be generated from those regions 
(e.g. Alpar \& Shaham 1985; Lamb et al. 1985).
It was only with the RXTE discovery  of the twin $\sim 1$~kHz QPOs from neutron star
systems that signals in the range of dynamical
frequencies of the innermost disk could finally be studied (Strohmayer et al. 1996;
van der Klis et al. 1996).
 
Most QPO models were introduced in an effort to interpret the properties
of neutron star QPOs. In particular a generic model soon emerged in which 
the faster kHz signal arises from the orbital (i.e. azimuthal)
motion of matter in the inner disk region around the neutron star,
as Sunyaev (1972) and others had suggested in the early seventies. 
The same idea was applied to the faster QPOs from black holes, soon
after their discovery (Morgan, Remillard \& Greiner 1997). 
The phenomenology of QPOs in black hole systems 
is not as rich as that of neutron star binaries; however black holes provide a 
{\it cleaner} environment to probe matter motion in the very strong 
gravitational field regime, as they do not possess a solid surface, a 
stably-anchored magnetic field, nor a boundary layer at the disk inner
edge. 

Several models that
were originally developed to interpret neutron star QPOs are not
applicable to black hole QPOs, as they require the presence of 
a star surface and/or (offset) magnetic field. 
Among these are beat frequency models (Lamb et al. 1985;
Miller, Lamb \& Psaltis 1998). 
Other models are applicable to both neutron star and black
hole QPOs and build on the analogy between the QPO modes and continuum 
power spectral components that are 
observed in both types of systems, especially the QPO pair  
at higher frequencies. Among these are the relativistic procession
model and the epicyclic resonance model. We emphasize that both 
models are aimed at explaining the observed QPO frequencies
based on the idea that QPOs are produced at a given radius in the
disk: they are thus {\it local} models. We discuss them briefly below
and emphasise that they provide also an alternative way of 
estimating black spin (methods based on X-ray spectrum 
continuum-fitting and reflection/Fe-line modelling are described 
elsewhere in this volume).

\subsection{The Relativistic Precession Model}
\label{sec:RPM}

\begin{figure*}
\includegraphics[width=1.0\textwidth]{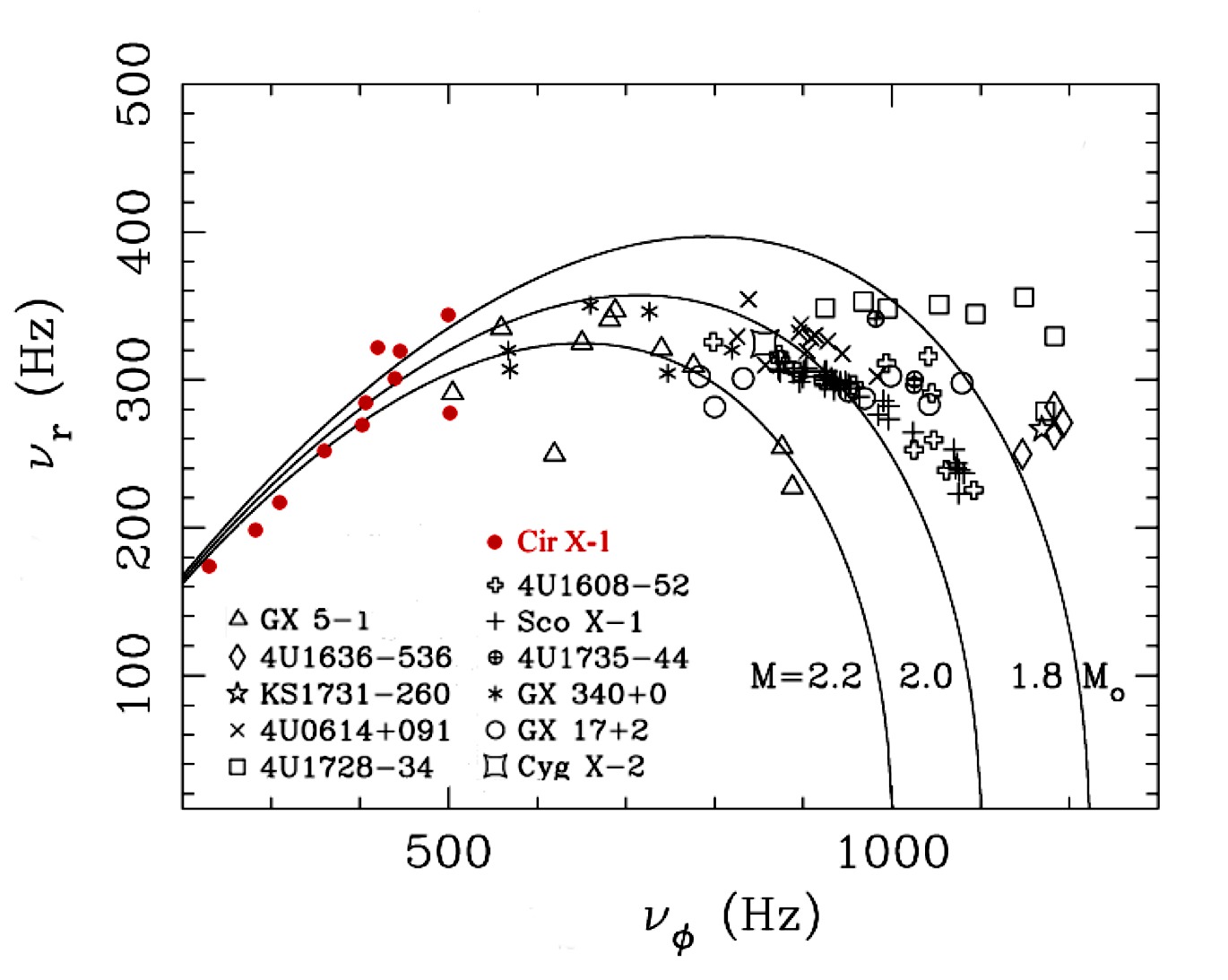}
\caption{kHz QPO frequency difference, versus upper kHz QPO frequency for a 
number of LMXRBs. Error bars are not plotted. 
Since in the RPM the upper kHz QPO corresponds to $\nu_\phi$ and the 
lower kHz QPO to $\nu_{per}$, their difference $\nu_\phi-\nu_{per}$ 
represents the radial precession frequency: therefore the curves 
give the $\nu_r$ vs. $\nu_\phi$ of matter in nearly circular orbit 
around a non-rotating neutron star, of mass 2.2, 2.0 and 1.8~$M_\odot$ 
(a Schwarzschild spacetime is used here). 
The filled red circles refer to Cir X-1; these data 
provide a confirmation of the low frequency behaviour of kHz QPO 
predicted by the RPM (after Stella \& Vietri 1999; Boutloukos et al. 2006)}
\label{fig:nu_r_vs_nu_phi}      
\end{figure*}

The Relativistic Precession Model (RPM) was originally aimed 
at interpreting the twin kHz QPOs as well as a low-frequency
QPO mode (the HBOs) of neutron star of low mass X-ray binaries
of both the Atoll and Z-classes (Stella \& Vietri 1998, 1999).
In the RPM the higher and lower frequency kHz QPOs are identified
respectively with the azimuthal frequency $\nu_\phi$ and the periastron
precession frequency $\nu_{per}$ of matter orbiting
at a given radius in innermost disk regions. In terms of the fundamental
frequencies of motion, the latter frequency is $\nu_{per} = \nu_\phi-\nu_r$,
such that the difference frequency of the kHz QPOs,
corresponds to the radial epicyclic frequency $\nu_r$. 
Fig. \ref{fig:nu_r_vs_nu_phi} shows different curves for the way in which $\nu_{r}$ 
is expected to vary as a function of increasing $\nu_{\phi}$, while the radius 
at which the QPOs are produced decreases. The space-time around 
neutron stars is approximated here by the equations in Fig. \ref{fig:freqs} 
with $a=0$, i.e. a Schwarzschild spacetime. (The effects
induced by neutron star rotation on the $\nu_r - \nu_{\phi}$ 
relation are small, though non-negligible, see Stella \& Vietri 1999). 
Plotted curves 
are for selected values of the neutron star mass.
The measured values of the difference frequency of the twin 
kHz QPO vs. the frequency of the upper kHz QPOs for 11 neutron star
LMXRBs is also plotted.
It is apparent that for masses of $\sim 2$~M$_{\odot}$,
the simple model above is in semi-quantitative agreement with measured
values, including the  decrease of the difference frequency for increasing 
QPO frequency, which was seen  
in Sco~X-1, 4U~1608-52, 4U~1735-44 and 4U~1728-34. Note that most 
measured points lie near the broad peak region of the radial epicyclic frequency.   
The behaviour of $\nu_r$ at low frequencies
is confirmed by the QPO separation in Cir~X-1 
which was found to increase with QPO frequency (Boutlokous et al. 2006).

In the RPM the HBO frequency is related to the nodal 
precession frequency, $\nu_{nod}$,  at the same radius 
where the signals at $\nu_\phi$ and $\nu_{per}$ are produced.
$\nu_{nod}$ is given by  $\nu_\phi - \nu_\theta$, the difference
between the azimuthal frequency and the vertical epicyclic
frequency. In the weak field limit, the relativistic nodal precession,
scales as the square of the azimuthal frequency, 
as in the well-known Lense-Thirring effect. This 
dependence is in agreement with that observed in a number of 
neutron star systems (Stella \& Vietri 1998, Ford et al. 1998; 
Psaltis et al. 1999). 

The RPM reproduces accurately also 
the PBK correlation, without resorting to additional assumptions.
The measured QPO and peaked-noise frequencies giving rise to the PBK
correlation is given in Fig.~1B of Stella, Vietri \& Morsink (1999), 
together with the curves obtained 
for $\nu_{nod}$ and $\nu_\phi$ as a function of $\nu_{per}$ for 
some selected cases cases of rotating neutron star models.
The RPM dependence of $\nu_{nod}$ on $\nu_{per}$
matches nicely the observed correlation 
over $\sim 3$ decades in frequency and encompasses both neutron
star and black hole systems.
Indeed this was the context in which the RPM was first 
applied to black hole QPOs and broad noise components.
 
Figure 1A in Stella, Vietri \& Morsink (1999) 
shows frequency measurements for 
black hole systems, together with 
$\nu_{nod}$ and $\nu_\phi$ 
as a function of $\nu_{per}$ for
selected values of the mass and spin parameter.
$\nu_{nod}$
depends weakly on $M$ and strongly on $a$; the opposite holds 
for $\nu_\phi$. In principle, black hole mass and angular momentum
can be thus be measured if three QPO frequencies 
(type C QPO, together with two high frequency QPOs) 
are detected simultaneously in a system. 
For the range of masses encompassed 
by black holes in X-ray binaries (roughly $5-10$~M$_\odot$) 
the application of the RPM to black hole QPOs indicates
relatively small values of $a/M \sim 0.1-0.3$. 

\begin{figure*}
\includegraphics[width=1.0\textwidth]{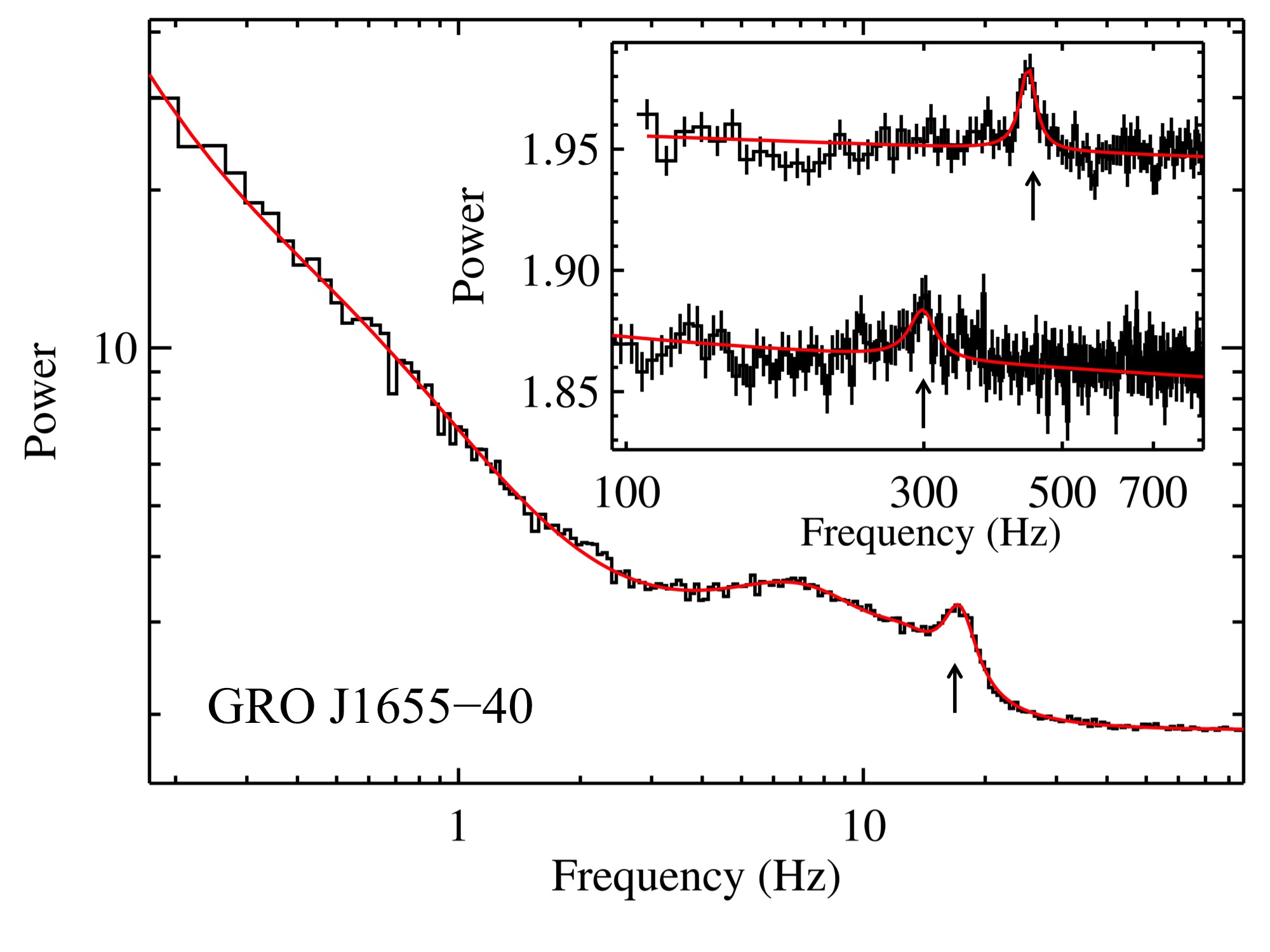}
\caption{Power spectrum of GRO J1655-40 displaying three simultaneous 
QPO peaks (marked by arrows): the type C at $\sim18$~Hz, upper and lower high frequency 
QPO at $\sim 300$ and $\sim450$~Hz, respectively (after Motta et al. 2014a).}
\label{fig:PSD_1655}       
\end{figure*}

\begin{figure*}
\includegraphics[width=1.0\textwidth]{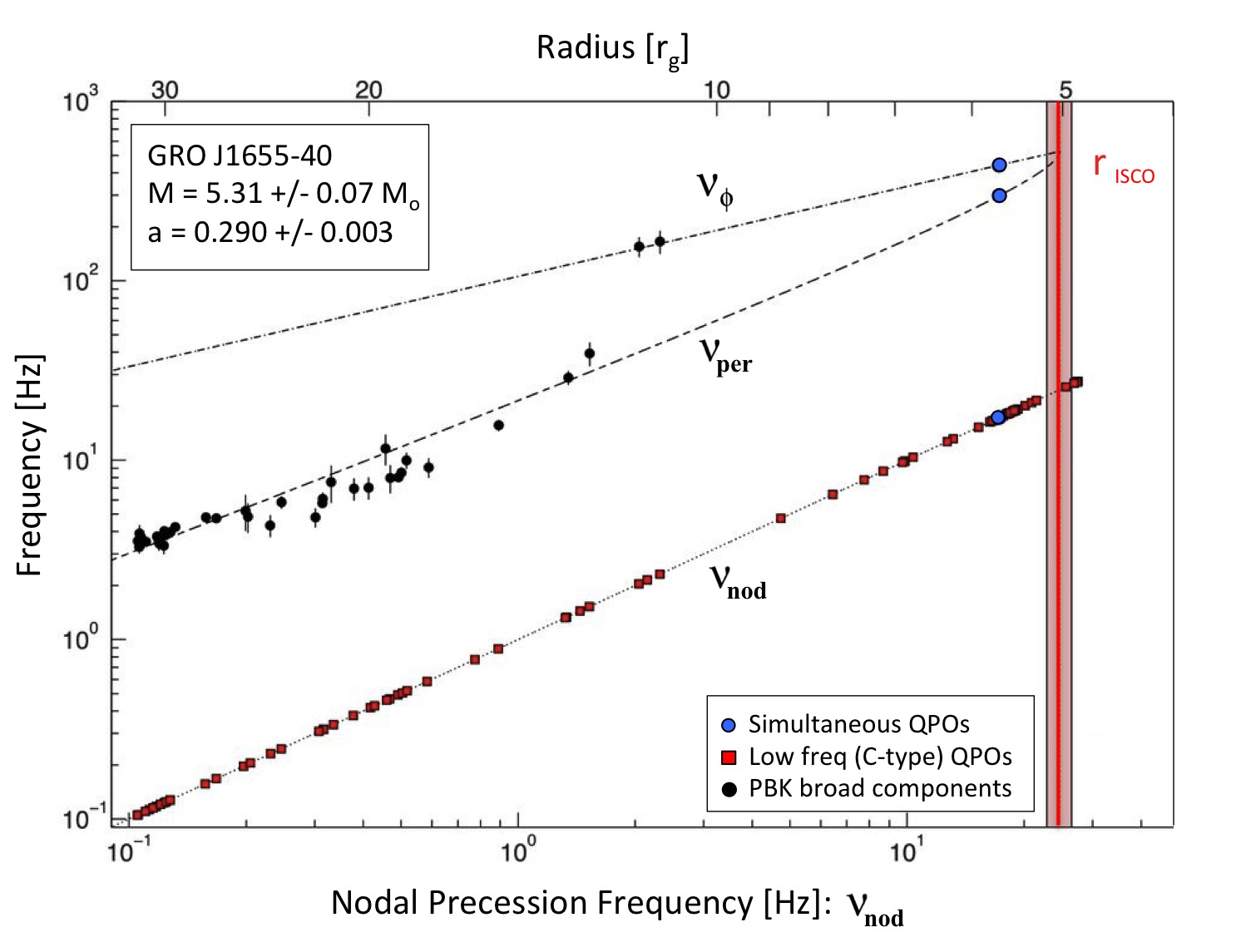}
\caption{Nodal precession frequency (dotted line), periastron precession frequency (dashed line) and orbital frequency (dotÐdashed line) as a function of the nodal precession frequency around a Kerr black hole as predicted by the RPM. The lines are drawn for the mass and spin values that provide fit best fit the three simultaneous QPOs observed in GRO J1655-40 (blue points in the plot). The corresponding radii are given in the top x-axis. The black circles are the characteristic frequencies of the source's broad power spectrum components; they all lie close to the low-frequency extrapolation of the frequencies predicted by the RPM, as derived on the three simultaneous points only. The squares, giving the frequency of type C QPOs plotted against itself, illustrate the range of frequencies (and thus radii) over which these QPOs are seen. The red line marks the ISCO radius and nodal frequency; and the red vertical band indicates its corresponding $3\sigma$ uncertainty (after Motta et al. 2014a).}
\label{fig:QPOs_1655}      
\end{figure*}

\begin{figure*}
\includegraphics[width=1.0\textwidth]{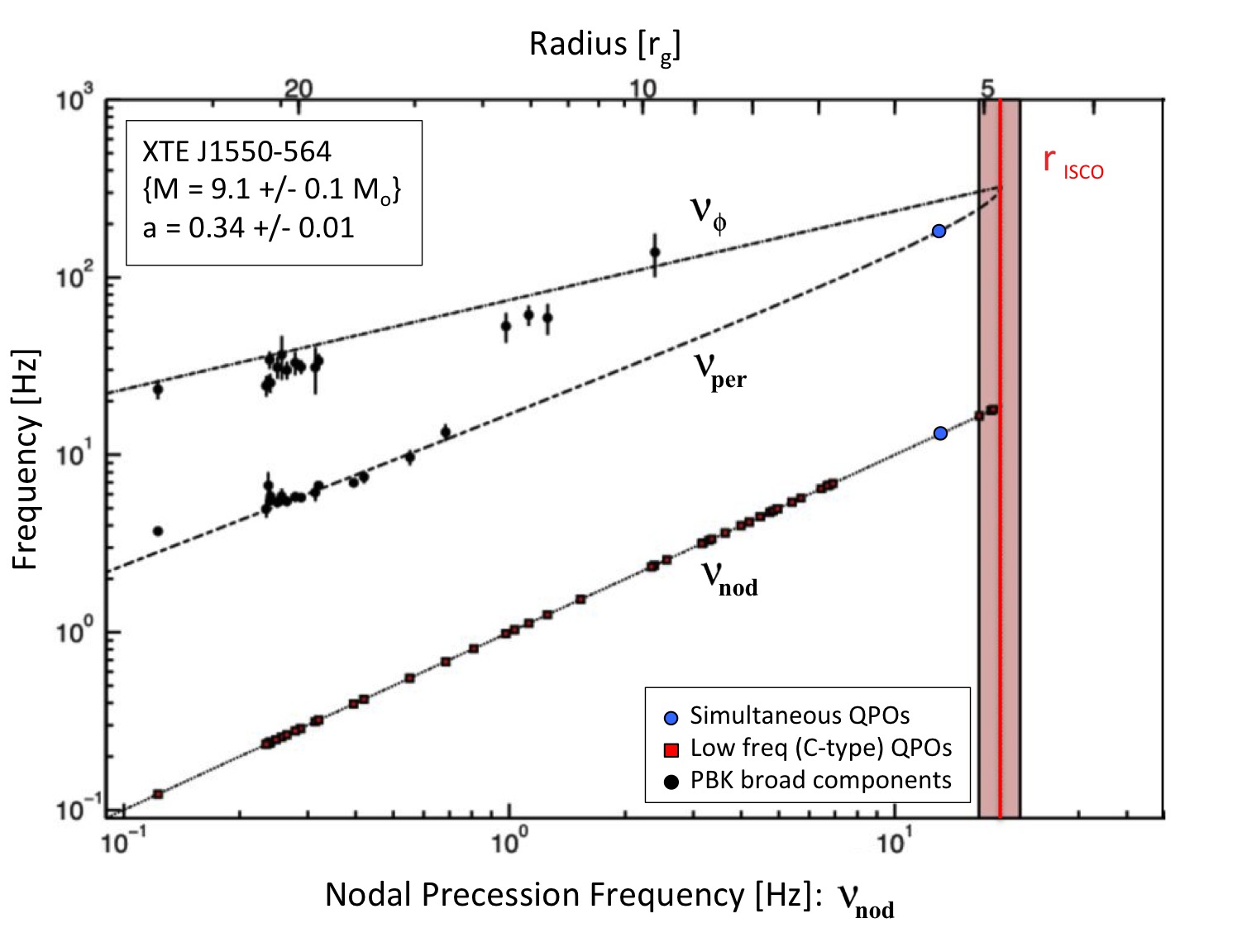}
\caption{Same as Fig.~\ref{fig:QPOs_1655}, but the points here are from XTE J1550-564. Note that in this case only 
two simultaneous QPOs (blue points in the plot) were observed, a type C QPO and the lower high frequency QPO; the best fit was obtained by using the black hole mass derived from 
optical spectrophotometric measurements (after Motta et al. 2014b).}
\label{fig:QPOs_1550}      
\end{figure*}

Only in one instance a type C low-frequency QPO was detected
simultaneously with two high-frequency QPOs in a black
hole system, GRO J1655-40 (see Fig.~\ref{fig:PSD_1655}). 
This made the first complete application
of the RPM to a black hole possible (Motta et al. 2014a).
By fitting the three measured QPO frequencies, precise values of the 
black hole mass (5.31 $\pm$ 0.07 M$_\odot$, consistent with the mass 
measured from optical/NIR spectro-photometric observations) and spin 
(a = 0.290 $\pm$ 0.003) were obtained through the sole use of X-ray timing.
Not only the quasi periodic oscillations, but also the broad band noise components 
of GRO J1655-40 and their variations match accurately the predictions 
of the RPM: this is shown in Fig.~\ref{fig:QPOs_1655}. 
The relative width of the QPO peaks was 
found consistent with being due to the jitter of the radius where
QPOs are generated.
In the case of XTE J1550-564 only type C and lower high-frequency QPOs
were observed simultaneously (Fig.~\ref{fig:QPOs_1550}). By using the black hole mass 
as determined from optical/NIR measurements, the RPM yields a value 
of the spin parameter of a = 0.34 $\pm$ 0.01 (Motta et al. 2014b).

\subsection{Epicyclic Resonance Models}
\label{sec:resonance}

Another family of {\it local} models exploits the fact that at some specific 
radii in the accretion disk, the radial $\nu_r$ and vertical $\nu_\theta$
epicyclic frequencies attain simple integer ratios and thus may resonate 
(Kluzniak \& Abramowicz 2001, 2002; Abramowicz \& Kluzniak 2001).
Parametric and forced resonances have been discussed, especially
those with small integer ratios 2:1, 3:1, 3:2. 
Examples of two such resonances are shown in Fig.~\ref{fig:freqs}. 
Resonances between $\nu_r$ and the azimuthal frequency $\nu_\phi$ have
also been considered (T\"or\"ok et al. 2005). These Epicyclic Resonance Models 
(ERMs) successfully explain black hole QPOs 
with frequency ratio consistent with 2:3 or 1:2 (see Sect. 3.2). 
As a given resonance condition is verified only at a fixed radius 
in the disk, the QPO frequencies are expected to remain constant, 
or jump from one resonance to another. 
In the application to neutron star QPOs various commensurabilities 
of epicyclic frequencies with other frequencies (e.g. with the spin
frequency) have also been considered, which can help interpret 
kHz QPOs in the framework of ERMs. This included also the development
of some {\it tunable} versions of the model, which can produce 
fairly continuum variations of the kHz QPO frequencies around the 
resonance condition. 
Simple applications of ERMs to black hole high frequency QPO pairs 
indicate large values of the spin parameter ($a > 0.9$) in the case 
of the 3:2 parametric resonance, and somewhat smaller values 
($a \sim 0.3 - 0.6$) for forced resonances (T\"or\"ok et al. 2005). 
In ERMs low frequency QPOs (type C and HBOs in particular) 
remain to be interpreted. 

\subsection{Disk Oscillation Models}
\label{sec:DOM}

Different types of disk oscillation modes have been 
considered as mechanisms that can give rise to QPOs and/or 
rapid noise-like variability in accreting black hole systems.
 
Inertial $g$-modes can be trapped in the resonant 
cavity that is created by the broad peak of the radial
epicyclic frequency (see Fig.~\ref{fig:freqs}). 
They give rise to the same global oscillations over a range of
disk radii, the fundamental frequency of which is close to 
the maximum value of $\nu_r$ (Nowak \& Wagoner 1991, 1992; 
Nowak et al. 1997). Global corrugation modes are expected to
occur in disks around spinning compact objects because of
the nodal precession induced by frame dragging 
(Silbergleit et al. 2001; Wagoner et al. 2001). 
Frame dragging can drive also warping modes close 
to the disk inner boundary at the ISCO (Markovic \& Lamb 
1998; Armitage \& Natarajan 1999). In disk $p$-modes the 
restoring force arises from pressure gradients;  
the fundamental mode in a small region close to ISCO 
attains higher frequencies than the local value of 
$\nu_r$ (Wagoner et al. 2001). 
In a series of papers Kato studied trapped non-axisymmetric 
$g$-mode and vertical $p$-mode oscillations that are 
resonantly excited in the inner disk regions; 
mode frequencies for
both magnetic and non-magnetic disks are compared
with QPO frequencies from black holes as well as 
neutron star systems (Kato 2012a,b,c and references therein;
see also Li, Goodman \& Narayan 2003).
The global stability of non-axisymmetric trapped $p-$modes 
was analyzed by Lai \& Tsang (2009) who showed that modes 
with frequencies $\sim (0.5 Ð 0.7) m \nu_r(ISCO)$
(with $m = 1, 2, 3 ... $ is the azimuthal wavenumber)
can grow as a result of corotational instability. 
Some of their models have the largest growth rate 
for $ m=2$ and 3, which might explain the 3:2 frequency ratio
of some black hole high frequency QPOs.

\subsection{Other Models}
\label{sec:sims} 

A variety of other models has been proposed which involve at 
least one fundamental frequency of motion in the strong-field regime.  
Ingram, Done \& Fragile (2009) consider a model in which the Lense-Thirring 
effect drives solid body-like nodal precession of a radially extended region 
of the hot inner flow. With the aid of numerical simulation they estimate the 
inner radius of the precessing region, work out the precession 
frequency and apply it to the low frequency QPOs of black holes (see also 
Ingram \& Done 2012 and references therein). 
Petri (2008) studied the oscillations caused by the parametric resonance 
induced by the interaction between a spiral density wave in 
the accretion disk, excited close to the ISCO, and vertical 
epicyclic oscillations.

Tagger \& Varniere (2006) developed the 
so-called accretionÐejection instability model for discs threaded by 
large-scale and intense poloidal magnetic fields. 
This B-field provides a strong coupling at corotation 
between spiral density waves and Rossby waves, leading to
amplification and injection of energy to a hot disk corona.

There are also models in which the oscillation frequency is
not related to the fundamental frequencies of motion. 
Among these is the disk-jet oscillation model, in which black 
hole-threated field lines
interact with the inner disk, extracting rotational energy and powering
jet ejection through the Blandford-Znajek mechanism; 
oscillations at 1/4, 1/2, and 3/4 the rotation frequency at the 
black hole horizon are especially prominent in the 
general relativistic  magnetohydrodinamical (MHD) simulations of 
magnetically-choked accretion flows by McKinney et al. (2012), the
latter two being in a 3:2 frequency ratio. 

\subsection{QPOs in accretion disk simulations}
\label{sec:sims} 

Searches for diskseismic modes were carried out in both 
hydrodynamical and MHD disk
models. In the hydrodynamical simulations of 
Reynolds \& Miller (2009) and O'Neill, Reynold \& Miller  
(2009) trapped g-modes and inner p-modes in a narrow range 
of frequencies close to the maximum radial epicyclic frequency
were found. On the contrary in the MHD case, 
turbulence created by the magneto-rotational 
instability (MRI) did not excite 
disk seismic modes, epicyclic frequencies, nor resonances with
3:2 frequency ratio at a level that could be revealed by 
simulations. On the other hand MRI turbulence was found
to give rise to local hydrodynamic waves.

In the GR MHD disk simulations 
of Wellons et al. (2014) fluctuations induced 
by MRI-driven MHD turbulence were 
followed over a fairly long time and synthetic
light curves of the disk emission as seen by a distant 
observed were calculated, by taking into account 
relativistic effects in the disk. 
The corresponding power spectra show a power-law like 
continuum and broad feature at high frequency, whose amplitude 
increases with observer inclination. Such a 
high-frequency feature is a product of the 
relativistic Doppler effect
in the innermost disk regions and encodes information
on the black hole mass and spin. 

Despite the progress achieved in recent years, 
much remains to be learned from 
the study of disk variability and QPOs in disk 
simulations (see e.g. Blaes 2014; Fragile 2014; 
Henisey et al. 2009; Dolence et al. 2012; 
Romanova et al. 2013).
The inclusion of full radiative transfer in 3D
disk simulations will be especially 
important for the generation synthetic light
curve simulations that can be directly compared to
the observations. Steps in this direction 
have been undertaken (Jiang et al. 2013;
Sadowski et al. 2014). 

\subsection{Conclusion}
\label{sec:concl} 

The past years (1996-2012) have seen an impressive increase of phenomenology of
fast variations from BHBs, mostly due to the RXTE satellite. The available
observational picture is very complex, but clear patterns have emerged and
theoretical models to interpret them are now available and have been tested. 
Despite the amount of information available however, the number of detections
of high-frequency oscillations from BHBs is very limited, due to their faintness and
elusiveness. This is in contrast to the strong signals observed from neutron-star 
binaries, where the same mechanisms for QPO production could be at
work. As the RXTE mission has been terminated, the observational picture can be improved
only with new missions. The ASTROSAT satellite, expected to be launched in 2015,
will provide an invaluable contribution by re-opening the window onto fast timing variability.
Its sensitivity below 10 keV is similar to that of RXTE, but above that energy the effective
area of its LAXPC instrument is considerably larger, offering a good opportunity to 
detect high-frequency oscillations. In addition, capitalizing on the RXTE heritage, 
it will be possible to concentrate observations in the outburst intervals when these oscillations
have been detected, maximizing the probability of increasing the number of detections.

Most of what is currently known (and briefly surveyed in the present 
review) on the fast variability of BHBs relies upon X-ray timing information
with little energy resolution, like that afforded by proportional 
counters (e.g. RXTE's). 
The potential of energy-resolved studies capable of addressing 
spectral variability and delays on dynamical timescales is beginning
to emerge from observations of supermassive black holes in active galactic nuclei
with CCD-type energy resolution (Iwasawa et al. 2004; Zoghbi et al. 2012). 
Relativistically shifted and broadened 
Fe K$\alpha$ lines from the innermost disk regions and their fast variability 
in combination with the fast variability of the X-ray continuum spectral components 
appear to be very promising in this context. In particular tomography and reverberation 
studies at X-ray energies in stellar mass accreting black holes will be 
able to achieve outstanding sensitivity to the dynamics of matter motion 
in the strong field regime (Uttley et al. 2014). To this aim very large area 
instrumentation ($\sim$5-10 m$^2$ in the classical 1-10 keV energy range) 
with CCD-like spectral resolution, such as that of the proposed 
mission LOFT (see: http://sci.esa.int/loft/53447-loft-yellow-book/\#), 
will be necessary.


\begin{acknowledgements}
We thank Sara Motta for insightful discussions. TMB acknowledges support from PRIN INAF 2012-6 ``Accreting X-ray binaries: understanding physics through periodic and aperiodic variability". LS acknowledges support from PRIN-INAF-2011 ``A deep insight into strong gravity around black-holes: new theoretical and observational approaches"

\end{acknowledgements}



\end{document}